# Signal Sets on Time Scales with Application to Hybrid Systems

Ti-Chung Lee, *Senior Member IEEE*, Ying Tan, *Senior Member IEEE*, Iven Mareels, *Fellow*, *IEEE*

*Abstract*—Recently, time scales calculus is developed to unify continuous and discrete analysis. By extending the definition of time scales properly, this paper introduces the concept of a signal set as well as its stability properties in terms of the so-called pseudo distance measure. This leads to more general Lyapunov like conditions to check stability properties of systems with hybrid nature. By way of examples, the proposed framework is used to model hybrid systems with simplicity and flexibility to characterize trajectories in the behavior of hybrid systems.

## I. INTRODUCTION

As engineering systems are getting more and more complex, they require both continuous and discrete valued signals, as well as continuous and discrete dynamics or time scales to describe their behaviors. Such systems have a hybrid nature. Examples include impulsive behavior in biological systems, or systems with logic rules and switching, control behavior induced through sampler-and-hold structures, engineering systems with impact forces and so on, see, for example [4, 8, 10, 20, 25] and references therein.

Recently, dynamic equations on time scales have been introduced to unify continuous-time and discrete-time dynamics [1, 3]. With the help of unified time scales calculus, the time scale framework can be used to investigate the properties of complex engineering systems [10, 12, 21, 24]. For example, it was pointed in [10] that a possible approach to model a hybrid system is to consider "dynamical systems on time scales". As a time scale is an arbitrary, but, fixed closed subset of the real line, it is not equivalent to the hybrid time domain defined in [10]. In this work, a novel concept of generalized time scales is introduced to fully capture the behavior of general engineered hybrid systems, including their behavior as time progresses indefinitely. This allows us to consider notions of stability.

With the help of a pseudo distance measure, the convergence properties of signal sets defined on generalized time scales are described with corresponding uniform global stability (UGS) and uniform global asymptotic stability (UGAS). Consequently, Lyapunov function-based conditions are obtained to verify UGS and UGAS. Two examples illustrate how the proposed framework works.

A special case of systems with the hybrid nature is hybrid systems. Different characterizations of hybrid systems have been proposed, see, for example, [6, 10, 14, 20, 24] and references therein. Compared with the widely used hybrid time domain (see [10]), there are three key advantages to represent hybrid systems using the notion of signal sets defined on generalized time scales: (1) The simplicity, as generalized time scales are just subsets of one dimensional space $\Re$. (2) The time scale calculus has been well developed, see [1, 3, 5]. (3) The flexibility. For example, switched systems and Zeno solutions can be handled using the proposed framework as shown in Section IV. C and Section IV. D.

This paper is an extended version of our conference paper [18] with the focus on introducing signal sets defined on generalized time scales. A more complete set of stability conditions is provided, and we add more examples to illustrate the flexibility of the framework, including a new modeling technique for switched systems.

The structure of this paper is as follows: Section II recalls some basic properties and defines two important operators relative to generalized time scales. Section III presents the general concept of signal sets defined on generalized time scales. Two important stability concepts (UGS and UGAS) and some Lyapunov based conditions are discussed. Section IV proposes a new definition of hybrid solutions such that it can be applied to the model studied in [10] as well as a model of arbitrarily switched systems. An example illustrates how the proposed framework can be used to describe the behavior of passing through a Zeno time. Conclusions are drawn in Section V.

**Notations**

1. $\Re = (-\infty, \infty)$, $\Re_+ = [0, \infty)$, $Z_+ = \{0, 1, \cdots\}$ and $\aleph = \{1, 2, \cdots\}$.
2. $\|v\|$ denotes the Euclidean norm of a vector $v \in \Re^p$, $p \in \aleph$.
3. For any nonempty subset $\Omega$ of $\Re^p$, let $\|v\|_\Omega = \inf_{u \in \Omega} \|u - v\|$ denote the distance function of a vector $v \in \Re^p$ w. r. t. $\Omega$.
4. With $a \le b$, $[a, b)$ is called a half-open interval.
5. A class $K_\infty$ function is a function $\beta : [0, \infty) \to [0, \infty)$ that is continuous, zero at zero and strictly increasing with $\beta(s) \to \infty$ as $s \to \infty$.

T. C. Lee is with Department of Electrical Engineering, Minghsin University of Science and Technology, Hsin-Hsing Road, Hsin-Fong, Hsinchu, Taiwan 304, R.O.C. (e-mail: tc1120@ms19.hinet.net ). This work was supported by the MOST, Taiwan, R.O.C., under contract MOST 108-2221-E-159-001-MY3.
Y. Tan is with Department of Mechanical Engineering, the University of Melbourne, VIC 3010, Australia (e-mail: yingt@unimelb.edu.au).
I. Mareels is Lab Director for IBM Research Australia (e-mail: imareels@au1.ibm.com), and Honorary Professor in the University of Melbourne (e-mail: i.mareels@unimelb.edu.au).

6. For a finite set $S$, $\#S$ denotes the total number of the elements of $S$. Moreover, $\#S = \infty$ when $S$ is an infinite set.
7. For any set $S \subseteq \Re^p$, $P(S)$ denotes the set of all subsets of $S$ and $rS = \{rs \mid s \in S\}, \forall r \in \Re$.

## II. PRELIMINARIES

Time scales are introduced in order to unify continuous and discrete analysis with many mathematical tools developed to facilitate such an analysis [1, 3, 5]. In this paper, the concept of time scales is generalized to describe the properties of systems with continuous and discrete signals over a half-open interval of the real numbers.

This section introduces generalized time scales with some of their properties and an interesting decomposition of the time parameter in terms of two operators related to continuous-time and discrete-time natures.

### A. Generalized time scales

The following definition describes the jumps in time (discrete) and some basic concepts in time scales.

**Definition 1.** Let $S$ be any subset of $\Re$.
a) Denote $Ini(S) = \inf_{t \in S} t$ and $Fin(S) = \sup_{t \in S} t$.
b) The function $\sigma : t \in S \mapsto \inf_{s > t, s \in S} s$ is said to be the forward jump operator where by the conventional usage, if $Fin(S) \in S$, let $\sigma(Fin(S)) = Fin(S)$.
c) Denote $R(S) = \{t \in S \mid \sigma(t) > t\}$.
d) For any $a, b \in \Re$,
$$S_{\leq a} = \{t \leq a \mid t \in S\} \text{ and } [a,b]_S = \{a \leq t \leq b \mid t \in S\}.$$

Definitions of time scale and generalized time scale are as follows.

**Definition 2.** a) A closed subset of $\Re$ is called a time scale.
b) A subset $I$ of $\Re$ is said to be a generalized time scale when $I_{\leq a}$ is a time scale for each $a \in I$. It is said to be forward (backward) complete when $Fin(I) = \infty$ ($Ini(I) = -\infty$). When $Fin(I) = \infty$ and $Ini(I) = -\infty$, it is said to be complete.

**Remark 1.** The concept of generalized time scale tries to extend the definition of the time scale to deal with a half-open interval $[a,b)$. Although this is not a time-scale, $[a,b)_{\leq c} = [a,c]$ are all time scale (closed) for all $a \leq c < b$. Such a generalization will be useful to characterize the solutions of hybrid systems. For example, the hybrid time domain in [10] includes a half-open interval. ∎

The following result directly follows from Definition 2. The proof is omitted.

**Lemma 1.** For a generalized time scale $I$, the following hold.
a) If $Ini(I) \in \Re$, $Ini(I) \in I$.
b) For any $a, b \in I$, $[a,b]_I$ is a time scale.
c) If $Fin(I) \in I$, $I$ is a time scale. Otherwise, every increasing sequence $\{t_n\} \subseteq I$ with $\lim_{n \to \infty} t_n = Fin(I)$ satisfies $I = \bigcup_{n=1}^{\infty} I_{\leq t_n}$. ∎

**Remark 2.** Roughly speaking, a) indicates that if the initial point exists (i.e., the solution starts from some time instant), it is in $I$. This is important when defining trajectories of dynamical systems because we need to specify initial conditions. In many instances, it may suffice to consider closed subintervals of a generalized time scale to analyze properties of a solution. Result b) indicates that any finite domain of interest can be assumed to be a time scale. Result c) states that a generalized time scale is just a set-theoretic limit of a sequence of "increasing" time scales with the same "initial point". This in itself provides another way of introducing generalized time scales, see [18]. ∎

Next the concept of subintervals is also needed. It will be used in the extension of solutions.

**Definition 3.** Let $I$ be a generalized time scale. A subset $J \subseteq I$ is said to be a subinterval of $I$ if $J$ is a generalized time scale and $[s,t]_J = [s,t]_I, \forall s, t \in J$.

### B. Two important operators

As time scales can represent both continuous and discrete solutions, two important operators are used to partition the time parameter into a continuous-time part and a discrete-time part, respectively.

**Definition 4.** Let $I$ be a generalized time scale with $Ini(I) \in I$. For any $t \in I$,
$$T_c^I(t) = Ini(I) + |[Ini(I),t]_I| \geq Ini(I) \quad (1)$$
is called the continuous-time part of $t$ relative to $I$. Moreover,
$$N_d^I(t) = \sum_{s \in R(I), s < t} [\sigma(s) - s] \geq 0 \quad (2)$$
is said to be the discrete-time part of $t$ relative to $I$. When the reference to $I$ is obvious from the context, we use the short-hand notation $t_c$ and $t_d$ to replace $T_c^I(t)$ and $N_d^I(t)$, respectively.

**Remark 3.** Notice that $|[Ini(I),t]_I|$ is the "occupying area" of $I$ contained in $[Ini(I),t]$. On the other hand, $t_d$ is "the other area" of the interval $[Ini(I),t]$ outside $I$. It is clear that the equality $t = t_c + t_d$ holds, as shown in Proposition 1 below, see also Fig. 1. This goes a long way to explain why $t_c$ and $t_d$ are said to be the continuous-time and discrete-time parts of $t$, respectively. They provide important information relative to the representation of different "time scales". ∎

$$Ini(I) \underset{\ell_1}{\rule{1cm}{0.4pt}} \underset{\mu_1}{\cdots\bullet} \underset{\mu_2}{\cdots\bullet} \underset{\mu_3}{\cdots\bullet} \underset{\ell_2}{\rule{0.5cm}{0.4pt}} t \quad I$$
$$t_c = Ini(I) + \ell_1 + \ell_2, \, t_d = \mu_1 + \mu_2 + \mu_3$$

Fig. 1: Illustrating $t_c$ and $t_d$.

It can be shown that the following basic properties hold (the proofs may be found in Appendix A).

**Proposition 1.** Let $I$ be a generalized time scale with $Ini(I) \in I$.
a) $T_c^I : I \to [Ini(I), \infty)$ is increasing and continuous with $T_c^I(Ini(I)) = Ini(I)$ and $T_c^I(I)$ being a closed or half-open interval.
b) For any $t \in I$, $t = T_c^I(t) + N_d^I(t)$. Particularly, $t \geq T_c^I(t)$.
c) $N_d^I : I \to \Re_+$ is increasing and continuous with $N_c^I(Ini(I)) = 0$. ∎

## C. Derivatives and absolute continuity on time scales

The general derivatives on time scales are defined following the similar concepts in [1, 3]:

**Definition 5.** Let $I$ be a time scale and $x: I \to \Re^p$ be a signal. With $t \in I$, $x^\Delta(t)$ is a delta (Hilger) derivative if, for any $\varepsilon > 0$ there exists $\delta > 0$ such that

$$\|x(\sigma(t)) - x(s) - x^\Delta(t)(\sigma(t) - s)\| \le \varepsilon \|\sigma(t) - s\| \quad (3)$$

for any $s \in I$ with $\|s - t\| < \delta$. When $\sigma(t) = t$, the notation
$$\dot{x}(t) = \lim_{s \to t, s \in I}(x(s) - x(t))/(s - t) \quad (4)$$
is also used to denote $x^\Delta(t)$.

**Remark 4.** Similar to the standard derivative in continuous-time functions, $x^\Delta(t)$ is unique. Roughly speaking, $x^\Delta(t)$ equals the differentiation of $x$ at $t$ in case of $\sigma(t) = t$ and equals the weighted difference $[x(\sigma(t)) - x(t)]/[\sigma(t) - t]$ of $x$ at $t$ in case of $\sigma(t) > t$. ∎

The concept of absolute continuity is important to define solutions of dynamical systems over time scales [5].

**Definition 6.** For a bounded time scale $I$, let $x: I \to \Re^p$ be a function. It is said to be absolutely continuous if, for any $\varepsilon > 0$ there exists $\delta > 0$ such that every finite pairwise disjoint family $\{I \cap [a_k, b_k]\}_{k=1}^n$ of subintervals of $I$ with $a_k, b_k \in I$ and $\sum_{k=1}^n (b_k - a_k) < \delta$, we have
$$\sum_{k=1}^n (x(b_k) - x(a_k)) < \varepsilon.$$

For a time scale $I$, a possible extension of a function $x: I \to \Re^p$ to $[Ini(I), Fin(I)]$ is defined as follows

$$\widetilde{x}(t) = \begin{cases} x(t), t \in I \\ x(s) + \dfrac{x(\sigma(s)) - x(s)}{\sigma(s) - s}(t - s), t \in (s, \sigma(s)) \text{ with } s \in I. \end{cases} \quad (5)$$

Lemma 2 shows that by using the extension, the absolute continuity (Definition 6) is the same as the continuous-time signals [5]. This leads to the fundamental theorem of calculus, see the discussions in e.g. [5, 15, 18].

**Lemma 2.** Let $I$ be a bounded time scale. Then, $x: I \to \Re^p$ is absolutely continuous if and only if $\widetilde{x}$ is absolutely continuous. Moreover, if $x$ is absolutely continuous, there exists a Lebesgue measure zero set $E \subseteq I \setminus R(I)$ such that $\dot{\widetilde{x}}(t) = x^\Delta(t)$ for all $t \in I \setminus E$. ∎

## III. SIGNALS DEFINED ON GENERALIZED TIME SCALES

This section introduces the concept of signal set defined on generalized time scales. As in [19], pseudo distance measure is employed to measure the convergence. Uniform global stability (UGS) and uniform global asymptotic stability (UGAS) are then defined. Some Lyapunov function-based conditions are proposed to check UGS and UGAS.

### A. General signal sets: Some stability concept

The concepts of signal set and pseudo distance measure are defined as follows:

**Definition 7.** Let $X$ be any non-empty set.
a) For a generalized time scale $I$, a function $x: I \to X$ is said to be a signal on $X$. For easy reference, denote $Dom(x) = I$.
b) A family $\Sigma$ consisting of signals on $X$ is said to be a signal set on $X$. The set of all signals on $\Re_+$ is denoted as P.
c) An operator $\Delta: \Sigma \to P$ is called as a pseudo distance measure when $Dom(x) = Dom(\Delta(x))$, $\forall x \in \Sigma$. For any $x \in \Sigma$, $\Delta(x)$ is called a pseudo distance function associated with $x$. It is then a function from $Dom(x)$ to $\Re_+$.

With the distance measure, the stability properties of signal sets defined on generalized time scales are described.

**Definition 8.** Let $\Sigma$ be a signal set on a non-empty set $X$ and $\Delta$ a pseudo distance measure. Let $d_x = \Delta(x), \forall x \in \Sigma$.

a) It is said that $\Sigma$ is *uniformly globally stable* (UGS) w.r.t. $\Delta$ if, there exists a class $K_\infty$ function $\beta$ such that for all $x \in \Sigma$, $d_x(t) \le \beta(d_x(s)), \forall t, s \in Dom(x)$ with $t \ge s$.

b) It is said that $\Sigma$ is *uniformly globally pre-asymptotically stable* (p-UGAS) w.r.t. $\Delta$ provided that the following hold:
1) $\Sigma$ is UGS w.r.t. $\Delta$.
2) For any $0 < \varepsilon < 1$, there exists $T(\varepsilon) > 0$ such that for any $x \in \Sigma$, if $d_x(s) \le 1/\varepsilon$, then $d_x(t) < \varepsilon, \forall t, s \in Dom(x)$ with $t \ge s + T$.

In addition that for any $x \in \Sigma$, there exists a forward complete $\widetilde{x} \in \Sigma$ such that $Dom(x)$ is a subinterval of $Dom(\widetilde{x})$ and $\widetilde{x}(t) = x(t), \forall t \in Dom(x)$, it is said that $\Sigma$ is *uniformly globally asymptotically stable* (UGAS) w.r.t. $\Delta$.

**Remark 5.** Condition 2) is called uniform global attractivity. Usually, this property is defined using two arbitrarily chosen parameters $\varepsilon > 0$ and $r > 0$ with the constant $T(\varepsilon)$ and $1/\varepsilon$ being replaced by $T'(\varepsilon, r)$ and $r$, respectively. Here $r = 1/\varepsilon$ is used for simplification. It is not difficult to check that these two definitions are in fact equivalent. ∎

**Remark 6.** The p-UGAS property was considered in [25] to highlight that solutions of interests may only be defined on some bounded subsets of $\Re$, which can happen in hybrid systems. When all solutions can be extended to forward complete solutions, p-UGAS becomes the standard UGAS property defined for continuous-time dynamic systems. ∎

For any signal $x \in \Sigma$, $d_x = d \circ x$ is the associated distance function where $d: X \to [0, \infty)$ is a distance function. Such an associated distance function can be used as a pseudo distance measure as shown in Fig 2. When $X \subseteq \Re^p$ and $d = \|\cdot\|$ is the Euclidean norm, the property of UGAS is related to the asymptotic convergence of the origin.

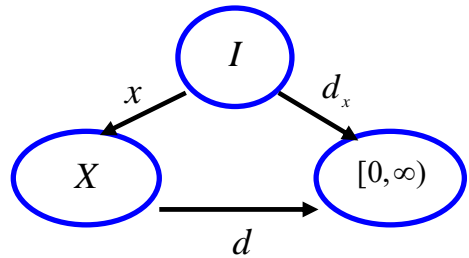

Fig. 2: A possible pseudo distance measure.

The following example illustrates the need of using signal set.

**Example 1.** Consider the following system [13]:
$$x_1^\Delta = -x_1 + x_2^2$$
$$x_2^\Delta = -x_2 - x_1 x_2. \quad (6)$$
For any solution $x = (x_1, x_2)^T : I \to \Re^2$, consider the pseudo distance function $d_x(t) = \|x(t)\|, \forall t \in I$. To study the stability property of solutions, let $V_x(t) = x_1^2(t) + x_2^2(t) = d_x^2(t), \forall t \in I$, be another associated distance (Lyapunov) function.

Based on the calculation in [13], we have
$$V_x^\Delta(t) = -2V_x(t) + (\sigma(t) - t)V_x(t)(1 + x_2^2(t)), \forall t \in I. \quad (7)$$
When $I = \Re_+$, $\sigma(t) = t$ and $\dot V_x = V_x^\Delta = -2V_x$. Let $\Sigma_c$ denote the set of all solutions defined on $\Re_+$. This implies $d_x(t) = e^{-(t-s)}d_x(s), \forall 0 \le s \le t$. Thus, $\Sigma_c$ is UGAS w.r.t. $\Delta_c$ where $\Delta_c : x \in \Sigma_c \mapsto d_x$ is a distance measure.

When $I = rZ_+, r > 0$, we have $\sigma(t) = t + r$ and
$$V_x(t+r) = V_x(t) + rV_x^\Delta(t) = V_x(t)((r-1)^2 + r^2 x_2^2(t)), \forall t \in I.$$
Let $\Sigma_r$ denote the set of all solutions defined on $rZ_+$. If $r < 2$, only local stability can be guaranteed. When $r > 2$, the origin is unstable. Particularly, $\Sigma_r$ is not UGS w.r.t. $\Delta_r$ where $\Delta_r : x \in \Sigma_r \mapsto d_x$ is a distance measure.

Three different situations of stability then appear for different kinds of solutions.

**Remark 7.** This example shows that different classes of solutions may have different stability properties. This indicates that signal sets can be used to represent partial solutions of a system with their stability properties. Moreover, this example illustrates the unification of time scales as the continuous-time solutions and the discrete-time solutions can be treated using the same framework. ∎

### B. Lyapunov like conditions

To guarantee the UGS and UGAS properties, the Lyapunov approach is used, with so called weak Lyapunov functions. The main stability result is captured in Theorem 1 (its proof can be found in Appendix B).

**Definition 9.** Let $\Sigma$ be a signal set on a non-empty set $X$ and $\Delta$ a pseudo distance measure. An operator $V : \Sigma \to P$ is said to be a *K-weak Lyapunov functional of* $\Sigma$ w.r.t. $\Delta$ if it is a pseudo distance measure, and there exist three class-$K_\infty$ functions $\alpha, \beta$ and $\gamma$ such that for any $x \in \Sigma$, the following statements hold where $d_x = \Delta(x)$ and $V_x = V(x)$:
1) $\alpha(d_x(t)) \le V_x(t) \le \beta(d_x(t)), \forall t \in Dom(x)$.
2) $V_x(t) \le \gamma(V_x(s)), \forall t, s \in Dom(x)$ with $t \ge s$.
When $\gamma$ is the identity function, $V$ is said to be a *weak Lyapunov functional of* $\Sigma$ w.r.t. $\Delta$.

**Theorem 1.** Let $\Sigma$ be a signal set on a non-empty set $X$ and $\Delta$ a pseudo distance measure. Then, the following hold.
a) When there exists a *K-weak Lyapunov functional* $V$ of $\Sigma$ w.r.t. $\Delta$, $\Sigma$ is UGS w.r.t. $\Delta$.
b) The signal set $\Sigma$ is p-UGAS w.r.t. $\Delta$ if and only if $\Sigma$ is UGS w.r.t. $\Delta$ and the following holds with $d_x = \Delta(x)$:
**(C1)** For any $\varepsilon > 0$, there exists $T(\varepsilon) > 0$ such that there are no $x \in \Sigma$ and $s, t \in Dom(x)$, with $t \ge s + T$ and such that

$$\varepsilon \le d_x(\tau) \le 1/\varepsilon, \forall \tau \in [s,t]_{Dom(x)}. \quad (8) \quad ∎$$

Corollary 1 follows from Theorem 1 (the proof is given in Appendix C). It can be used to derive a Krasovskii-LaSalle like theorem when an appropriate compactness property of the signal set holds.

**Corollary 1.** Let $\Sigma$ be a signal set on a non-empty set $X$ and $\Delta$ a pseudo distance measure. If there exists a weak Lyapunov functional $V$ of $\Sigma$ w.r.t. $\Delta$, then $\Sigma$ is UGS w.r.t. $\Delta$. Moreover, $\Sigma$ is p-UGAS w.r.t. $\Delta$ under the following extra conditions where $d_x = \Delta(x)$ and $V_x = V(x)$: :
1) There exists $M > 0$ such that
$$\sigma(t) \le t + M, \forall t \in Dom(x), \forall x \in \Sigma.$$
2) For any $\varepsilon > 0$, there exist $T(\varepsilon) > 0$ and $\delta(\varepsilon) > 0$ such that for any $x \in \Sigma$ and $s, t \in Dom(x)$ with $t \ge s + T$,
$$\varepsilon \le d_x(\tau) \le 1/\varepsilon, \forall \tau \in [s,t]_{Dom(x)} \Rightarrow V_x(t) \le V_x(s) - \delta. \quad ∎$$

**Remark 8.** Particularly, $\Sigma$ is p-UGAS w.r.t. $\Delta$ when $V$ is strict, i.e., it is a pseudo distance measure and there exist class-$K_\infty$ functions $\alpha, \beta$ and $\gamma$ such that for any $x \in \Sigma$, the first condition of Definition 9 holds and $V_x$ is absolutely continuous on $[s,t]_I, \forall s, t \in I$, such that $V_x^\Delta(\tau) \le -\gamma(d_x(\tau))$ for any $\tau \in I \setminus E$ where $I = Dom(x)$ and $E \subseteq I \setminus R(I)$ is a Lebesgue measure zero set. ∎

Example 2 illustrates the use of Theorem 1.

**Example 2.** Consider the following switched systems:
$$\Sigma_0 : \begin{aligned} \dot x_1 &= 10 x_2 \\ \dot x_2 &= 0 \end{aligned} \quad \Sigma_1 : \begin{aligned} \dot x_1 &= 1.5 x_1 + 2 x_2 \\ \dot x_2 &= -2 x_1 - 0.5 x_2. \end{aligned} \quad (9)$$
Notice that both subsystems are unstable. It was used in [8] to illustrate that under a suitable switching law, the switched system's trivial solution is stable. Following [16] and using Theorem 1, we demonstrate how a *K*-weak Lyapunov functional is enough to conclude that the trivial solution is UGAS.

Indeed, as in [8], consider the switching law
$$\sigma(t) = \begin{cases} 0, \text{if } (x_1 + 4x_2)(x_1 - 2x_2) \le 0 \\ 1, \text{if } (x_1 + 4x_2)(x_1 - 2x_2) > 0. \end{cases} \quad (10)$$
Let $\Sigma$ be the set of all maximal solutions starting at $t_0 = 0$. For all $x = (x_1, x_2)^T \in \Sigma$, consider the Lyapunov like functional $V_x = 2(x_1(\cdot) + x_2(\cdot))^2 + x_2^2(\cdot)$. To investigate the convergence of the origin, we consider the pseudo distance function $d_x = \|x(\cdot)\|$. It can be checked that the first condition of Definition 9 holds with $\alpha(s) = s^2/3, \beta(s) = 5s^2, \forall s \ge 0$.

As discussed in [16], the switched system is running clockwise with the "angular velocity" satisfying
$$-10 \le [\tan^{-1}(x_2(t)/x_1(t))]' \le -1/2, \forall t \ge 0.$$
Let the plane be divided into the following two regions:
$$X_0 : (x_1 + 4x_2)(x_1 - 2x_2) \le 0; \; X_1 : (x_1 + 4x_2)(x_1 - 2x_2) \ge 0.$$
Notice that $X_0$ and $X_1$ consist of two circular sectors with infinite radius. Moreover, the switched system is switched to the subsystem $\Sigma_i$ on mode $i$ when $(x_1, x_2)^T \in X_i$, $i = 0, 1$. Thus, any solution (except the trivial solution) stays in $X_i$ with dwell time $\tau_{min} \leq \tau_i \leq \tau_{max}$ where $\tau_{min}$ and $\tau_{max}$ are two fixed positive constants. Thus, $x$ is forward complete.

On the mode $i = 1$, $|x_1| \geq 2|x_2|$ and hence
$$\dot{V}_x\big|_{\Sigma_1} = -2x_1^2 + 5x_2^2 \leq -(x_1^2 + x_2^2)/2, \forall (x_1, x_2)^T \in X_1. \quad (11)$$
So $V$ is a weak Lyapunov functional. However, on the mod $i = 0$, $V$ is not a weak Lyapunov functional. But, it is indeed a "$K$-weak" Lyapunov functional as shown below.

Notice that both subsystems are linear. Thus, there exist $\rho \geq 1$ and $M > 0$ such that
$$V_x(t) \leq \rho e^{M|t-s|} V_x(s), \forall s, t \in \mathfrak{R}_+. \quad (12)$$
If $0 \leq s \leq t \leq s + 2\tau_{max}$, the second condition of Definition 9 holds with $\gamma(\tau) = \rho^2 e^{2M\tau_{max}} \tau, \forall \tau \geq 0$. When $t > s + 2\tau_{max}$, let $s'$ and $t'$ be the first and the last time instants, respectively, that the solution stays in $X_1$ on the time interval $[s,t]$. Thus, $s \leq s' \leq s + \tau_{max}$ and $t - \tau_{max} \leq t' \leq t$. For any time instant $l$ that the switched system switches from the mode $i = 1$ to $i = 0$, let $l' \geq l$ be the first time instant that the solution leaves $X_0$ (and enters $X_1$). By the second equation of the subsystem $\Sigma_0$, $x_2(l) = x_2(l')$ and
$$V_x(l) = V_x(l') \quad (13)$$
due to $|x_1 + x_2|_{x_1 = 2x_2} = |3x_2| = |-3x_2| = |x_1 + x_2|_{x_1 = -4x_2}$. This, together with (11), gives us that $V_x(t') \leq V_x(s')$. Hence
$$V_x(t) \leq \rho e^{M\tau_{max}} V_x(t') \leq \rho e^{M\tau_{max}} V_x(s') \leq \rho^2 e^{2M\tau_{max}} V_x(s).$$
So the second condition of Definition 9 also holds. Therefore $V$ is a $K$-weak Lyapunov functional.

To show UGAS, it remains to check (C1). For any $\varepsilon > 0$, let $T = 2(k+1)\tau_{max}$ with $k > 1 + 10/(\varepsilon^4 \tau_{min})$ being a positive integer. If there exists $x \in \Sigma$ satisfying (8) for some $s, t \in Dom(x)$, we will find a contradiction. Let $s'$ and $t'$ be defined as above. Since $t \geq s + T$, $t' \geq s' + 2k\tau_{max}$. For each $1 \leq n \leq k$, let $t_0 = s'$, and $t_{2n-1} \in [t_{2n-2}, t']$ be the first time instant that the switched system switches from the mode $i = 1$ to $i = 0$ and $t_{2n} \in [t_{2n-1}, t']$ the first time instant that the solution leaves $X_0$ (and enters $X_1$ again). Thus,
$$0 \leq V_x(t_{2k}) \leq V_x(s') + \sum_{1 \leq n \leq k} \int_{t_{2n}}^{t_{2n+1}} \dot{V}\big|_{\Sigma_1}(\tau) d\tau$$
$$\leq 1/\varepsilon^2 [5 - (k-1)\tau_{min}\varepsilon^4 / 2] < 0 \quad (14)$$
in view of the choice of $k$, (11) and (13). We reach a contradiction. Hence (C1) holds and $\Sigma$ is UGAS w.r.t. $\Delta: x \in \Sigma \mapsto d_x$. By Theorem 1, the origin is then UGAS.

## IV. APPLICATIONS TO HYBRID SYSTEMS

In order to illustrate the usefulness of the proposed framework: signal sets on time scales, we use it to propose a general model for hybrid systems. In particular, we show that the concept of solutions defined in [10] is a special case of our proposed framework. Moreover, our framework makes it possible to model arbitrarily fast switched systems and bouncing balls (a hybrid system) with the consideration of passing through a Zeno time.

### A. Solutions of hybrid systems defined on generalized time scales

Hybrid systems defined on time scales are introduced first. For $C, D \subseteq \mathfrak{R}^p$, consider the following hybrid system:
$$\dot{x} \in F(x), x \in C \quad (15)$$
$$x^+ \in G(x), x \in D \quad (16)$$
where $F$ and $G$ are two set-valued functions from $\mathfrak{R}^p$ to $R^p$.

Definition 11 defines the solutions of (15)-(16) on generalized time scales. Here a locally absolutely continuous function $x$ means that it is absolutely continuous on $[s,t]_{Dom(x)}, \forall s,t \in Dom(x)$.

**Definition 11. (Solutions defined on generalized time scales)** For $I \subseteq \mathfrak{R}$ being a generalized time scale, a locally absolutely continuous function $x: I \to \mathfrak{R}^p$ is said to be a solution of (15)-(16) if the following conditions hold.
a) **(Initial condition)** $x(Ini(I)) \in C \cup D$ when $Ini(I) \in I$.
b) **(Flow condition)** For $t \in I \setminus (R(I) \cup \{Fin(I)\})$, $x(t) \in C$ and there exists a measure zero set $S$, with $R(I) \subseteq S \subseteq I$ and such that for any $t \in I \setminus S$, $\dot{x}(t)$ exists and $\dot{x}(t) \in F(x(t))$.
c) **(Jump condition)** For any $t \in R(I)$, $x(t) \in D$ and $x(\sigma(t)) \in G(x(t))$.

A solution $\tilde{x}$ of (15)-(16) is said to be an extension of $x$ if $Dom(x)$ is a subinterval of $Dom(\tilde{x})$ and $\tilde{x}(t) = x(t), \forall t \in Dom(x)$. A solution $x$ of (15)-(16) is said to be maximal if it has no extension. Throughout this paper, we assume that some solutions exist for (15)-(16).

### B. Hybrid solutions defined on hybrid time domains

Now we show that the concept of solutions for a hybrid system using the proposed framework includes the model proposed in [10] as a special case. Throughout this subsection, let $\mathrm{H} \subset P(\mathfrak{R})$ consist of generalized time scales $I$ with $Ini(I) = 0$ and $\sigma(t) = t + 1, \forall t \in R(I)$.

First, let us recall the definition of hybrid time domain proposed in [10]. Let $j \in Z_+ \cup \{\infty\}$ and $\{t_n\}_{n=0}^j \subseteq \mathfrak{R}_+$ be a finite or infinite increasing sequence with $t_0 = 0$. Any set of the form
$$\hat{I} = (\bigcup_{0 \leq i < j} [t_i, t_{i+1}] \times \{i\}) \cup (\hat{I}_j \times \{j\}) \quad (17)$$
is said to be a hybrid time domain where $\hat{I}_\infty$ is the empty set and in case of $j < \infty$, $\hat{I}_j$ has one of the following forms:
$$\hat{I}_j = [t_j, \infty), \hat{I}_j = [t_j, t_{j+1}), \quad (18)$$
where $t_{j+1} \in [t_j, \infty)$.

Consider the sum function $\ell: \mathfrak{R} \times \mathfrak{R} \to \mathfrak{R}$ defined by $\ell(s,t) = s + t, \forall s, t \in \mathfrak{R}$. The following result characterizes the relation between hybrid time domains and generalized time scales. Due to space limitation, the proof is omitted.
**Proposition 2.** The following statements hold.
a) For any non-empty hybrid time domain $\hat{I}$, $I = \ell(\hat{I}) \in \mathrm{H}$. Moreover, $T_c^I(t+i) = t$ and $N_d^I(t+i) = i$ for any $(t,i) \in \hat{I}$.
b) Conversely, for any $I \in \mathrm{H}$,
$$\hat{I} = \{(T_c^I(t), N_d^I(t)) \mid t \in I\} \quad (19)$$
is a non-empty hybrid time domain.

*Remark 9*. Proposition 2 implies that the non-empty hybrid time domains are one-to-one corresponding to those generalized time scales $I \in H$. Hence, hybrid time domains defined in [10] can be viewed as a special case of generalized time scales. ∎

For any hybrid solution $x: I \to \Re^p$ of (15)-(16) with $I \in H$, the function $\hat{x}: \hat{I} \to \Re^p$ defined as $\hat{x}(s,j) = x(s+j), \forall (s,j) \in \hat{I}$, is then a hybrid solution of (15)-(16) in the sense of [10] where $\hat{I}$ is defined as in (19). Conversely, let $\hat{x}: \hat{I} \to \Re^p$ be a hybrid solution of (15)-(16) in the sense of [10]. With $I = \ell(\hat{I})$, the function $x: I \to \Re^p$ defined as $x(t) = \hat{x}(T_c^I(t), N_d^I(t)), \forall t \in I$, becomes a hybrid solution of (15)-(16). In other words, by using the transformation of defining domains, one hybrid solution defined as in [10] is just a hybrid solution defined on a generalized time scale $I \in H$.

### C. Using hybrid systems to model switched systems

With $\Lambda = \{1, 2, \cdots, N\}$, consider the switched system
$$\dot{x} = f(x, \lambda) \quad (20)$$
where $x \in \Re^p$ is the state vector, $\lambda$ is a $\Lambda$-valued switching signal and $f: \chi \subseteq \Re^p \times \Lambda \to \Re^p$ is a function. For a precise definition of solutions, we refer readers to [17]. Let $x: I_* \to \Re^p$ and $\lambda: [0, \infty) \to \Re^p$ form a solution pair of (20) where $I_* = [0, a)$ with $a \in \Re_+ \cup \{\infty\}$. As usual, $\lambda$ is assumed to be a piecewise constant right-continuous function and has finitely many points of discontinuity in any finite time interval. Let $J$ consist of all discontinuous points of $\lambda$ and be described in the ascending order as follows:
$$J = \{t_1 < t_2 < \cdots < t_{\#J}\} \text{ or } J = \{t_1 < t_2 < \cdots\}.$$
For $0 < r \le 1$, define a function $S_J^r: \Re_+ \to P(\Re_+)$ as
$$S_J^r(s) = \begin{cases} \{s\}, \text{if } 0 \le s < t_1 \\ \{s + \sum_{m=1}^n r^m\}, \text{if } t_n < s < t_{n+1}, 1 \le n < \#J \\ \{s + \sum_{m=1}^{n-1} r^m, s + \sum_{m=1}^n r^m\}, \text{if } s = t_n, 1 \le n \le \#J \\ \{s + \sum_{m=1}^{\#J} r^m\}, \text{if } \#J < \infty \text{ and } s > t_{\#J}. \end{cases} \quad (21)$$
Notice that if $\#J = \infty$, $\lim_{n \to \infty} t_n = \infty$. Thus, (21) is well-defined. Let $I = \bigcup_{s \in I_*} S_J^r(s)$. It can be verified that $I$ is a generalized time scale and $T_c^I(t) = s, \forall s \in I_*, \forall t \in S_J^r(s)$.

Let $\tilde{x}: I \to \chi \subseteq \Re^{p+1}$ be defined as
$$\tilde{x}(t) = (x(t_c), \lambda(t_c)), \forall t \in I,$$
where $t_c = T_c^I(t)$. Let $C = D = \chi, F = (f^T, 0)^T$ and
$$G(u, \zeta) = \{(u, i) \in \chi | i \in \Lambda\}, \forall (u, \zeta) \in \chi.$$
Then, it can be seen that $\tilde{x}$ is a hybrid solution of (15)-(16).

In this way, all solutions of the switched system (20) can be transformed into hybrid solutions of a hybrid system. When one switched system has the so-called impulse effects, it can also be represented as a hybrid system by modifying the definition of the jump map $G$.

If we choose $r = 1$, the formulation of switched systems under the proposed framework is the same as the (two dimensional) hybrid time domain defined in [9, 23]. Under such a situation, it is difficult to model switched systems under arbitrarily switching without dwell-time condition. In particular, this type of formulation cannot well define uniform attractivity due to possibly infinite number of switching in a finite time interval for a family of switching signals.

However, our formulation still allows to model arbitrarily switched system by selecting $0 < r < 1$. It has
$$N_d^I(t) \le \sum_{m=1}^{\infty} r^m = r/(1-r), \forall t \in I. \quad (22)$$
Thus, $t \ge t_c = T_c^I(t) \ge t - r/(1-r), \forall t \in I$. Hence all time intervals with lengths, say $L$, will be transformed into generalized time scales with their upper bounds larger than or equal to $L$, and less than $L + r/(1-r)$. In this way, the proposed framework can model arbitrarily switched systems with well-defined uniform attractivity. This shows the flexibility of the proposed framework.

### D. Passing through zeno time: A simple example

Consider the following hybrid model of bouncing ball:
$$\dot{x} = \begin{bmatrix} 0 & 1 \\ 0 & 0 \end{bmatrix} x + \begin{bmatrix} 0 \\ -g \end{bmatrix}, x \in C = \Re_+ \times \Re \quad (23)$$
$$x^+ = -\theta x, x \in D = \{0\} \times (-\Re_+) \quad (24)$$
where $0 \le \theta < 1$ is a constant and $g$ is the gravity constant.

Let $t_0 = 0$ and $t_i, i \in \aleph$, are the $i$-th time instant of the ball hitting on the ground. For any given initial height $h_0 \ge 0$ and initial velocity $v_0 \ge 0$, it is easy to compute that
$$t_1 = (v_0 + \sqrt{v_0^2 + 2gh_0})/g \text{ and } v_1^- = -\sqrt{v_0^2 + 2gh_0}$$
where $v_1^-$ is the velocity of ball at $t = t_1$ before hitting the ground. Let $v_1^+ = -\theta v_1^-$ be the velocity of ball at $t = t_1$ after hitting the ground. It is easy to show that
$$t_{i+1} - t_i = 2v_i^+/g, v_{i+1}^- = -\theta v_{i+1}^- = \theta v_i^+, \forall i \in \aleph$$
where $v_i^+$ ($v_i^-$) is the velocity of ball at $t = t_i$ after (before) hitting the ground. This shows that $\{t_i\}_{i=1}^{\infty}$ is a geometry series with the common ratio $0 \le \theta < 1$. Thus, $t_\infty = \lim_{i \to \infty} t_i < \infty$. It is called a Zeno time because there are infinitely many switching before $t = t_\infty$.

By a careful check, we know that state (of height and velocity) will approach the zero as the time approaches $t_\infty$. Thus, it is possible to consider a solution $\tilde{x}$ that can be extended to $\Re_+$ with $\tilde{x} = (0,0)^T$ when passing through the Zeno time. How can we find such an extension?

Technically, we need to properly extend the defining domains so that the solutions can be still defined when passing the Zeno time [7]. Moreover, the extended solution may not be a solution of the original hybrid systems as indicated in [2, 22]. That suggests that one model might not be sufficient enough to capture a complete solution. Hence, more models are needed.

For the model described in (23)-(24), it is impossible to find a solution defined on $\Re_+$ with $\tilde{x}(t) = (0,0)^T, \forall t \ge t_\infty$.

Similar to the idea presented in [7], another model is generated using a modified equation of (23) as follows:

$$\dot{x} = \begin{bmatrix} 0 & 1 \\ 0 & 0 \end{bmatrix} x + \begin{bmatrix} 0 \\ -g \end{bmatrix} \eta(x), x \in \Re_+ \times \Re \quad (25)$$

where $\eta : \Re_+ \times \Re \to \Re$ is defined as follows:

$$\eta(x) = \begin{cases} 1, x \neq 0 \\ 0, x = 0. \end{cases} \quad (26)$$

Next it will show how the proposed time scale framework can provide a possible extended solution.

For $v_0 = h_0 = 0$, let $I = \bigcup_{n \in Z_+} \{1-(1/2)^n\} \cup [1,\infty)$. Otherwise, let $I = \bigcup_{s \in [0,t_\infty)} S_J^{1/2}(s) \cup [t_\infty + 1, \infty)$ where $S_J^{1/2}$ is the function defined as in (21) with $J = \{t_n | n \in \aleph\}$. Then, $I$ is a nonempty closed set. So $I$ is a time scale. Let

$$\tilde{x}(t) = \begin{cases} (h_0 + v_0 t - gt^2/2, v_0 - gt)^T, t \in [0,t_1] \\ (v_n^+ \bar{t} - g\bar{t}^2/2, v_n^+ - g\bar{t})^T, t \in [t_n + r_n, t_{n+1} + r_n] \\ (0,0)^T, t \geq t_\infty + 1 \end{cases}$$

where $n \in \aleph$, $r_n = \sum_{m=1}^{n} (1/2^m) = 1 - (1/2^n)$ and $\bar{t} = t_c - t_n = t - r_n - t_n$. It can be checked that $\tilde{x} : I \to \Re^2$ is locally absolutely continuous and is a solution of (24) and (25). Notice that the zeno time $t_\infty$ is converted to a point $t_\infty + 1 \in I$. Moreover, the real-time solution $x(s)(=\tilde{x}(t)$ with $s = t_c)$ can be described as follows:

$$x(s) = \begin{cases} \{(h_0 + v_0 s - gs^2/2, v_0 - gs)^T\}, s \in [0,t_1) \\ \{(v_i^+ \bar{s} - g\bar{s}^2/2, v_i^+ - g\bar{s})^T\}, s \in (t_n, t_{n+1}), n \in \aleph \\ \{(0, v_n^-)^T, (0, v_n^+)^T\}, s = t_n, n \in \aleph \\ \{(0,0)^T\}, s \in [t_\infty, \infty), \end{cases} \quad (27)$$

where $\bar{s} = s - t_n$. Thus, $\tilde{x}$ can be used to study the dynamic behavior of bouncing ball including the flow and jump phenomena, and the real-time solution (possibly, a set-valued function) $x$ to go back to the solution of the original system. This demonstrates another flexibility of the proposed framework.

## V. CONCLUSION

This work generalized the concept of time scale to describe solutions of hybrid systems. With the help of concept of signal sets, it is possible to characterize stability properties of signals using pseudo distance measure. Weaker Lyapunov functionals were introduced to verify uniform global stability and uniform global attractivity. The proposed concept of signal sets (defined on generalized time scales) was used to build a general model for hybrid systems. It was shown that the well-known model proposed in [10] is a special case of the proposed framework. Moreover, the proposed framework allows one to model systems that cannot be modeled easily using other approaches [10] such as fast switching systems and a bouncing ball system with Zeno solutions. Our future work focuses on extending the available analysis tools to study even more general systems.

## APPENDIX A: PROOF OF PROPOSITION 1

First, observe that

$$\left|[Ini(I),t]_I\right| = \int_{Ini(I)}^{t} \mu_I(\tau) d\tau$$

where $\mu_I$ is the indicator function, i.e., $\mu_I(\tau) = 1$ in case of $\tau \in I$, and $\mu_I(\tau) = 0$ in case of $\tau \notin I$. This indicates that $T_c^I$ is increasing and continuous with $T_c^I(Ini(I)) = Ini(I)$ [15]. By the intermediate value theorem, $T_c^I(I)$ is a closed or half-open interval contained in $[Ini(I), \infty)$. Hence a) holds.

Observe that

$$[Ini(I),t] = [Ini(I),t]_I \cup (\bigcup_{s<t, s \in R(I)} (s, \sigma(s))).$$

Hence

$$t - Ini(I) = \left|[Ini(I),t]\right| = \left|[Ini(I),t]_I\right| + \sum_{s \in R(I), s<t} (\sigma(s) - s).$$

This implies that $t = T_c^I(t) + N_d^I(t), \forall t \in I$. So b) holds.

By b), $N_d^I(t) = t - T_c^I(t)$ is continuous. In view of definition, $N_d^I$ is increasing with $N_d^I(Ini(I)) = 0$. This completes the proof of the proposition. ∎

## APPENDIX B: PROOF OF THEOREM 1

By definition, for all $x \in \Sigma$,

$$d_x(t) \leq \alpha^{-1}(V_x(t)) \leq \alpha^{-1}(\gamma(V_x(s))) \leq \alpha^{-1} \circ \gamma \circ \beta(d_x(s)), \quad (A1)$$

for any $t, s \in Dom(x)$ with $t \geq s$. Since $\alpha^{-1} \circ \gamma \circ \beta$ is also a class $K_\infty$ function, $\Sigma$ is UGS w.r.t. $\Delta$ in view of (A1). Result a) is then done.

By definition, p-UGAS implies UGS and (C1). Indeed, with the constant $T(\varepsilon) > 0$ given as in the second condition of p-UGAS, (8) cannot hold due to $d_x(s) \leq 1/\varepsilon$ and $d_x(t) < \varepsilon, \forall t, s \in Dom(x)$ with $t \geq s + T$. This shows the "only if" part of b). It remains to show the "if" part.

Suppose $\Sigma$ is not p-UGAS w.r.t. $\Delta$. Then, there exists $0 < \varepsilon_0 < 1$ such that for each $n \in \aleph$, there exist $x_n \in \Sigma$ and $s_n, t_n \in Dom(x_n)$ such that $t_n \geq s_n + n$, $d_{x_n}(s_n) \leq 1/\varepsilon_0$ and $d_{x_n}(t_n) \geq \varepsilon_0$. Let

$$\delta_0 = \min(\varepsilon_0, \beta^{-1} \circ \gamma^{-1} \circ \alpha(\varepsilon_0), 1/\alpha^{-1} \circ \gamma \circ \beta(1/\varepsilon_0)).$$

Then, $0 < \delta_0 < 1$. Under (C1), for any $n \geq T(\delta_0)$, we claim

$$\delta_0 \leq d_{x_n}(\tau) \leq 1/\delta_0, \forall \tau \in Dom(x_n) \text{ with } s_n \leq \tau \leq t_n. \quad (A2)$$

By (A1), for any $\tau \in Dom(x_n)$ with $\tau \geq s_n$,

$$d_x(\tau) \leq \alpha^{-1} \circ \gamma \circ \beta(d_{x_n}(s_n)) \leq \alpha^{-1} \circ \gamma \circ \beta(1/\varepsilon_0) \leq 1/\delta_0.$$

If there exists $s_n \leq \tau_0 \leq t_n$ such that $d_{x_n}(\tau_0) < \delta_0$, we have

$$\varepsilon_0 \leq d_{x_n}(t_n) \leq \alpha^{-1} \circ \gamma \circ \beta(d_{x_n}(\tau_0)) < \varepsilon_0$$

based on (A1). We reach a contradiction. Hence (A2) holds. For any $n \geq T(\delta_0)$, $t_n \geq s_n + n \geq s_n + T(\delta_0)$ and a $x_n \in \Sigma$ has been found such that (A2) (and (8)) holds. So a final contradiction is reached and $\Sigma$ is p-UGAS w.r.t. $\Delta$. This completes the proof of the theorem. ∎

## APPENDIX C: PROOF OF COROLLARY 1

According to Theorem 1, it remains to show (C1). For any $0 < \varepsilon < 1$, let $T(\varepsilon)$ be the constant given in (2). We claim that (C1) holds with $T(\varepsilon)$ replaced by $k(\varepsilon)(T(\varepsilon) + M)$ where $k(\varepsilon)$ is a positive integer satisfying $k(\varepsilon) > \beta(1/\varepsilon)/\delta(\varepsilon)$. If the claim is not true, then there exists $x_0 \in \Sigma$ and $s_0, t_0 \in Dom(x_0)$ such that $t_0 \geq s_0 + k(\varepsilon)(T(\varepsilon) + M)$ and

$$\varepsilon \leq d_{x_0}(\tau) \leq 1/\varepsilon, \forall \tau \in [s_0, t_0]_{Dom(x_0)}. \quad (A3)$$

Since $\sigma(t) \leq t + M, \forall t \in Dom(x_0)$ and $t_0 \geq s_0 + k(T + M)$, there exist $\{s_1 < s_2 < \cdots < s_k\} \subseteq [s_0, t_0]_{Dom(x_0)}$ such that

$$s_{i-1} + T \leq s_i \leq s_{i-1} + T + M, \forall 1 \leq i \leq k. \quad (A4)$$

By (2), (A3) and (A4), we have
$$V_{x_0}(s_i) - V_{x_0}(s_{i-1}) \leq -\delta, \forall 1 \leq i \leq k.$$

This results in
$$0 \leq V_{x_0}(s_k) \leq V_{x_0}(s_0) - k\delta \leq \beta(d_{x_0}(s_0)) - k\delta \leq \beta(1/\varepsilon) - k\delta < 0.$$

A contradiction is reached. Thus, the claim is true and (C1) holds. This completes the proof of the corollary. ∎